\documentclass[12pt]{article}
\usepackage{amsfonts,amsmath,amssymb,mathrsfs}

\title{The ``Unromantic Pictures'' of Quantum Theory}
\author{ 
Roderich Tumulka\footnote{Mathematisches Institut,
    Eberhard-Karls-Unversit\"at, Auf der Morgenstelle 10, 72076
    T\"ubingen, Germany.  E-mail:
    tumulka@everest.mathematik.uni-tuebingen.de}
}
\date{July 17, 2006}

\newcommand{\RRR}{\mathbb{R}}
\newcommand{\CCC}{\mathbb{C}}
\newcommand{\NNN}{\mathbb{N}}
\newcommand{\PPP}{\mathbb{P}}
\newcommand{\Hilbert}{\mathscr{H}}
\newcommand{\sphere}{\mathscr{S}}
\newcommand{\Q}{\mathcal{Q}}
\renewcommand{\sp}[2]{\langle #1 | #2 \rangle}
\newcommand{\vr}{r}%{\boldsymbol{r}}

\renewcommand{\Im}{\mathrm{Im}}
\newcommand{\foliation}{\mathscr{F}}
\newcommand{\M}{\mathscr{M}}
\newcommand{\cst}{{}^\sharp\! \M}%{\mathscr{C}}
\newcommand{\phys}{\mathrm{phys}}
\newcommand{\ext}{\mathrm{ext}}
\newcommand{\g}{{}^\sharp g}
\newcommand{\A}{{}^\sharp\! A}
\newcommand{\F}{{}^\sharp\! F}
\newcommand{\T}{{}^\sharp T}
\newcommand{\J}{{}^\sharp\! J}
\newcommand{\Foliation}{{}^\sharp\! \foliation}
\newcommand{\n}{{}^\sharp n}
\newcommand{\sgamma}{{}^\sharp \gamma}
\newcommand{\sepsilon}{{}^\sharp \varepsilon}
\newcommand{\sdelta}{{}^\sharp \delta}

\begin{document}\maketitle\sloppy
\begin{abstract}
I am concerned with two views of quantum mechanics that John S. Bell called ``unromantic'': spontaneous wave function collapse and Bohmian mechanics. I discuss some of their merits and report about recent progress concerning extensions to quantum field theory and relativity. In the last section, I speculate about an extension of Bohmian mechanics to quantum gravity.

\medskip

\end{abstract}
\tableofcontents

\section{On the Merits of the ``Unromantic Pictures''}

The quotation in the title is taken from a classic article of John S. Bell on ``Six possible worlds of quantum mechanics'' \cite{Bell86b}. He describes, discusses and comments on six views of quantum mechanics, three of which he calls ``romantic'': Complementarity, consciousness as the cause of wave function collapse, and the many-worlds view. And three he calls ``unromantic'': pragmatism, spontaneous wave function collapse, and Bohmian mechanics. This article is about the latter two views. These two unromantic pictures both fit into the category of ``quantum theories without observers'' (which also was a title of two conferences, in 1995 and 2004, and of three articles \cite{Pop,Bell86,Gol98}). In their unromantic attitude, they reject the idea that anything incomprehensible, or unanalyzable, or deep, or mysterious, or philosophical, is going on in a quantum measurement, and replace the vague talk usually surrounding the analysis of the measurement process by precise mathematics. And they dispense with observers in the sense that their fundamental formulations are not about the subjective experience of observers when making such-and-such experiments, but instead about (what they suggest as) objective physical reality.

Bohmian mechanics \cite{Bohm52,Bell66,survey,Gol01} takes the word ``particle'' literally and postulates that there are pointlike entities moving around in space, governed by an equation of motion, and thus have an actual and precise position at every time. These particles are the objective physical reality. 

Among the collapse theories, I will focus on the simplest (and perhaps best-known) one, the Ghirardi--Rimini--Weber (GRW) theory \cite{GRW86}. In this theory, the unitary Schr\"odinger evolution is replaced by a nonlinear, stochastic evolution for the wave function. Two versions of this theory are known, differing in their ontologies: according to ``GRWm,'' matter is continuously distributed, while according to ``GRWf,'' matter consists of discrete space-time points. 

Before I give the equations of these theories in Section 2, I will say in this section a few things about their relation to quantum mechanics. In Section 3 I will report recent progress concerning the extension of Bohmian mechanics and the GRW theory to quantum field theory and relativity. In Section 4 I close in a more speculative way, with a proposal for how to incorporate gravity into Bohmian mechanics.

\subsection{Quantum Mechanics Does Not \emph{Make} Predictions, It \emph{Is} the Prediction}

Observers are, in quantum theories without observers, not the protagonists of the axioms but merely particular physical systems governed by the same laws as any other physical system. The statement that observers will see this-and-this if they make such-and-such experiments then is a \emph{theorem}, not an axiom. For example, it has been shown \cite{Bohm52,Bell66,DGZ92, DGZ04} that observers in a typical Bohmian universe will see results of their experiments that appear random, with frequencies in agreement with the probability laws of quantum mechanics. 

This last statement is interesting. Put in other words, the formalism summarizing the predictions of Bohmian mechanics about observable effects agrees with the quantum formalism. Put succinctly, Bohmian mechanics implies the quantum formalism. If we regard, as it is often done, quantum mechanics as merely a set of rules for computing the possible outcomes of experiments and their probabilities, then quantum mechanics \emph{is} the prediction of Bohmian mechanics. Bohmian mechanics \emph{makes} predictions in the sense that it describes a model universe, for which the ``predictions'' are statements about what intelligent beings in that universe observe. Quantum mechanics, for comparison, does not \emph{make} predictions in this sense since it does not provide any model for the reality behind the appearances. For further comparison, the GRWm and GRWf theories also \emph{make} predictions, indeed identical predictions \cite{AGTZ06}, but predictions that differ from quantum mechanics. (The deviation is in most cases extremely small \cite{GRW86,Belljumps}, and the other cases are difficult to arrange. As a consequence, no experiment so far has been able to test the GRW theories against quantum mechanics \cite{BG03}.) The deviation underlines that the predictions do not \emph{have} to agree with quantum mechanics just because the theory involves a wave function.

The two unromantic pictures, Bohmian mechanics and spontaneous collapse, render it evident that quantum mechanics can be understood in terms of a completely coherent theory with a clear ontology. Regrettably, more than 50 years after Bohm and 20 years after GRW, this is still not very widely known. Given how vague and incoherent orthodox quantum philosophy is, and how radical the claims are that it makes about the intrinsic impossibility to understand physics, one might expect that scientists accept it only if they have to, under the load of incontrovertible evidence. One might thus expect that scientists would immediately give up orthodox quantum philosophy when they learn that theories exist that are understandable and account for all phenomena of quantum mechanics. But historically, the opposite was the case. When David Bohm argued in 1952, perhaps for the first time convincingly, that Bohmian mechanics accounts for quantum mechanics in terms of objective, but non-classical, particle trajectories, the reception was cold. What is it that motivates scientists, rational people who take pride in their ability to understand the most intricate theories, to give up on any serious understanding of quantum mechanics, in favor of the obscure orthodox quantum doctrine? I do not claim to be able to answer this question from the armchair. Indeed, I think that to determine the answer is a research topic for sociologists of science, and a worthwhile one. 

However, I would like to take you on a brief excursion in the following two subsections and consider two possible motivations.

\subsection{Positivism}

It is often taken as an objection against Bohmian mechanics that it entails the existence of unmeasurable quantities. For example, the velocity of a Bohmian particle cannot be measured if we do not know the wave function. \emph{What cannot, not even in principle, be measured,} I hear physicists say, \emph{cannot belong to a scientific theory. Rather, it is to be regarded like angels, ghosts, or the ether.} I would categorize this position as exaggerated positivism, and I find this argument surprising because quantum physicists should know first, of all scientists, that it is wrong.

The quantum formalism itself entails that nature can keep a secret, in the sense that there exist some facts that cannot be revealed by any experiment. To see how, we start from the mathematical fact that different ensembles of wave functions (mathematically represented by probability distributions $\mu$ over the unit sphere $\sphere(\Hilbert)$ of Hilbert space) can have the same density matrix $\hat\rho$, given by
\begin{equation}
  \hat\rho = \int\limits_{\sphere(\Hilbert)} |\psi \rangle \langle \psi| \, \mu(d\psi)\,.
\end{equation}
For example, an ensemble of spin-$\tfrac{1}{2}$ particles consisting of $50\%$ spin-up particles and $50\%$ spin-down has the same density matrix
\begin{equation}
  \hat\rho = \left( \begin{matrix} \tfrac{1}{2} & 0\\ 0 & \tfrac{1}{2} \end{matrix} \right)
\end{equation}
as the ensemble of $50\%$ spin-left and $50\%$ spin-right, or as the ensemble with the spin direction uniformly distributed over all directions. Since the statistics of any quantum experiment depends only on the density matrix, these different ensembles are \emph{empirically indistinguishable}. Nonetheless, they are physically different, since I may have prepared the spin state of every single member of the ensemble, so that I know the state vector of every particle. I can even prove that I know the state vector, and that nature remembers it, by naming, for every member of the ensemble, a direction and predicting with certainty the result of a Stern--Gerlach experiment for the spin component in this direction. Therefore, there is a matter of fact about whether the ensemble is an up--down ensemble or a left--right ensemble, but if I do not tell you, you have no way of determining which it is. It is a variant of this argument to say that \emph{one cannot measure the wave function} of an electron, even though there is, at least in some cases, a matter of fact about what its wave function is.

Once it is recognized that the position that I called exaggerated positivism is wrong, one realizes how eccentric it always was. It entails that every question can be answered by a suitable experiment, and that seems clearly wrong. Consider, for example, the question, ``who was Jack the Ripper?'' (This name was given to the unknown person who committed a series of brutal murders in London in the 1880s.) There must have been one or more persons who were guilty of these crimes, and several plausible suspects have been investigated, but as yet none could be proven guilty beyond reasonable doubt \cite{Pol}, and it seems quite possible that none ever will. So, questions need not be meaningless just because there is no systematic way of answering them. In the words of Bell \cite{Belljumps}: ``[T]o admit things not visible to the gross creatures that we are is, in my opinion, to show a decent humility, and not just a lamentable addiction to metaphysics.''

I want to mention another example against exaggerated positivism. The interior of a black hole is shielded from us by an event horizon, and since there is no way for us to learn about events inside the black hole, exaggerated positivism would imply that these events are not real. That sounds implausible. (It may sound even less plausible for de Sitter space-time \cite{HE73}, where large portions of space-time are shielded from each other by event horizons. There might be observers in both regions, and none of them has more right than the other to be regarded as ``outside.'') You may object that the events inside a black hole are \emph{in principle} observable, since if I get overwhelmed by curiosity I can cross the black hole's horizon, and then the desired information is accessible to me. But this is not quite true since there is a space-time region which remains inaccessible even if I enter the black hole because its causal future is disjoint from that of here-now. 

Moreover, there are subtleties about what it means for a quantity to be observable, which   are not easily appreciated if one is making axioms about observations, but become evident from the viewpoint of a quantum theory without observers. Let me illustrate them using again the example of the velocity in Bohmian mechanics: 
\begin{itemize}
\item If I know the wave function of a particle then I \emph{can} measure its velocity, but the experiment for this may change the particle's wave function completely.
\item If I know that the particle was, some time $\Delta t$ ago, within the radius $\varepsilon$ of the location $x\in \RRR^3$, then I \emph{can} measure the average velocity $\overline{v}$ over the time span $\Delta t$ by measuring the particle's present position $x' \in \RRR^3$ with inaccuracy $\varepsilon'$ and computing $\overline{v} = (x'-x)/\Delta t$ with inaccuracy $\leq (\varepsilon+\varepsilon')/\Delta t$. However, the velocity \emph{after} the detection may be quite different from $\overline{v}$, which was the average velocity \emph{before}.
\item As a consequence of the possibility of measuring average velocities, I \emph{can} measure \emph{asymptotic} velocities as long-term averages with arbitrarily high precision. Indeed, what is usually called a ``momentum measurement'' in quantum mechanics actually measures, in Bohmian mechanics, (mass times) the \emph{asymptotic} velocity in the absence of forces, but not (mass times) the \emph{instantaneous} velocity. 
\end{itemize}
This gives you an idea that the situation is more complex than conveyed the statement, ``velocities cannot be measured in Bohmian mechanics.''

Then what is the relevant difference between Bohmian velocities on the one hand and angels and the ether on the other that makes one scientific and the other two not, if both are empirically inaccessible? There are two differences. One is that there is no reason, if angels or the ether existed, why they should be unobservable. In Bohmian mechanics, in contrast, it is a consequence of the defining equations of the theory that velocities are unmeasurable. And the defining equations require nothing like fine tuning to entail this consequence. For comparison, a person who defends the existence of angels may easily end up postulating that angels do not want to be observed (not by skeptics, at least), a conspiratorial ad hoc postulate. The second difference concerns the following situation. If a theory claims the existence of some object $X$ in addition to matter, and if it turns out that $X$ has no influence on the behavior of matter, then one obtains a simpler theory by denying the existence of $X$ and keeping the same laws for matter. Such was the case with the ether. The key argument against the existence of the ether (or an absolute frame of reference) was not that Michelson and Morley could not observe it, but that Einstein could show how to obtain a theory that does not need the ether, by postulating a 4-dimensional reality (in space-time) instead of a 3-dimensional reality (in space) evolving with time. But such is not the case with the Bohmian velocities. If every particle has an actual position at every time, it necessarily has a velocity, so there is no way of keeping the positions without the velocities. And since, in Bohmian mechanics, matter \emph{consists} of the particles, removing the particles from the theory would remove the matter. Thus, the velocities do not form a superfluous superstructure like the ether.

\subsection{The Truman Show}

I turn to another reason that I think keeps many physicists from taking Bohmian mechanics seriously. They feel that \textit{a Bohmian universe, though it looks like a quantum universe, is not the real thing. Bohm is cheating in the sense that the outcome of a quantum measurement of an observable $A$ is not what one could regard as the true value of $A$, but rather just a random number with the right probability distribution, the one prescribed by the quantum formalism. Thus,} they conclude, \textit{the Bohmian world is a big fake like ``The Truman Show"---remember the movie with Jim Carrey \cite{truman}?---except that it is perfect. It is best regarded,} they feel, \textit{as something like a simulation of quantum mechanics, and the fact that there is no experiment that could distinguish Bohmian mechanics from quantum mechanics is no more relevant than the fact that you have no possibility to check experimentally whether you are a brain in the vat (fed by evil scientists with false sense data).}

Indeed, the result of, e.g., a Stern--Gerlach experiment in Bohmian mechanics cannot be regarded as the true value of the appropriate spin observable, but only as a random number with the right distribution. The same can be said of most experiments, the only exceptions being particle detectors, which do reveal the actual position of the Bohmian particle, and many experiments designed for measuring momentum, which do reveal (mass times) the asymptotic Bohmian velocity.

But why, in the first place, \emph{should} we believe in the idea that the result of a Stern--Gerlach experiment is ``the true value'' of the appropriate spin observable? Was this idea not already discredited by the ``two-valuedness'' of the result (i.e., by the fact that, for a spin-$\tfrac{1}{2}$ particle, only the two results $\pm \hbar/2$ are possible), in contradiction to the picture in which the result is a component of the (random) angular momentum vector? Was this idea, or more generally the one that \emph{quantum measurements} merely reveal pre-existing values of the \emph{observables}, not refuted in 1967 by the Kochen--Specker proof \cite{KS67} (and even earlier by Bell \cite{Bell66} and Gleason \cite{Gleason})? And was it not emphasized by orthodox quantum philosophy itself since the early days of quantum mechanics that the result is \emph{created}, rather than revealed, by the experiment? But then how could anybody ever expect quantum observables to have ``true values''? And how could anybody take seriously the objection that Bohmian mechanics is fake quantum mechanics? That is the real mystery.

To make a remark of a more sociological nature, I think it is the case that the orthodox view has a contradictory attitude towards the idea of ``true values,'' often (and misleadingly) called ``hidden variables.'' The typical orthodox physicist openly condemns hidden variables as impossible, but in his heart cannot abandon them, and continues to talk as if particles had energies and angular momentum vectors. Ironically, Bohmian mechanics is often called a ``hidden variables theory'' because it can be regarded, though somewhat inappropriately \cite{bac}, as postulating actual values for the position observable; but for all other observables, Bohmian mechanics does not claim the existence of hidden variables. Thus, in this sense Bohmian mechanics is more of a ``no-hidden-variables theory.'' The orthodox view seems much more obsessed with hidden variables than the Bohmian one, as it calls experiments ``measurements'' and operators ``observables,'' and as it regards the operator-observables as having roughly the same status as their namesakes energy, momentum, etc., in classical mechanics. In total, one easily tends to take operators too seriously, an attitude called ``naive realism about operators'' by Daumer \textit{et al.} \cite{DDGZ96,DGZ04}.

\subsection{It Is Matter That Matters}

A trait that distinguishes both Bohmian mechanics and the GRW theory from most other proposals about quantum reality is that they provide variables that represent the matter, more precisely that describe the distribution of matter in space and time. Such variables were called the ``primitive ontology'' by D\"urr \textit{et al.}\ \cite{DGZ92,DGZ04,AGTZ06} and the ``local beables'' by Bell \cite{Bellbook}. The simplest example of such variables are the (positions of the) particles in Bohmian mechanics. The GRW theory is known in two versions with different primitive ontologies, see Section \ref{sec:GRW}.

The attitude behind postulating such variables is to be contrasted with the attitude according to which the \emph{wave function} ``describes'' the state of the matter. The ``description'' provided by the wave function is, however, in such a vague sense that almost any two physicists disagree about what exactly the reality is like when the wave function is such-and-such. Note how different the sense is in which the ``primitive ontology'' provides a description of matter: If a theory postulates that matter consists of point particles, and provides the positions of these particles at all times, then it provides a picture that could not be sharper. It may be \emph{wrong}, but there is nothing \emph{vague} about it. But first of all, the primitive ontology makes \emph{explicit} what the reality is, rather than leaving it to everybody's private fantasies. This is a crucial merit of the unromantic pictures.

\section{Three Theories: Bohm, Collapse, and Again Collapse}

\subsection{Bohmian Mechanics}
\label{sec:bm}

Bohmian mechanics is a theory of (non-relativistic) particles in motion. The motion of a system of $N$ particles is provided by their world lines $t \mapsto Q_i(t)$, $i=1, \ldots, N$, where $Q_i(t)$ denotes the position in $\RRR^3$ of the $i$-th particle at time $t$. These world lines are determined by Bohm's law of motion \cite{Bohm52, Bell66, DGZ92, survey}, 
\begin{equation}\label{Bohm}
  \frac{dQ_i}{dt}=v_i^{\psi}(Q_1, \ldots, Q_N)=\frac{\hbar}{m_i}
  \Im \frac{\psi^{*}\nabla_i \psi}{\psi^{*}\psi}(Q_1\ldots,Q_N), 
\end{equation}
where $m_i$, $i=1, \ldots, N$,  are the masses of the particles;
the wave function $\psi$ evolves according to Schr\"odinger's equation
\begin{equation}\label{Schr}
 i\hbar\frac{\partial \psi}{\partial t} = H\psi \,,
\end{equation} 
where $H$ is the usual nonrelativistic Schr\"odinger Hamiltonian; for
spinless particles it is of the form
\begin{equation}
\label{eq:H}
H=-\sum_{k=1}^N\frac{\hbar^2}{2m_k}\nabla^2_k+V,
\end{equation}
containing as parameters the masses  of the particles as
well as the potential energy function $V$ of the system. 

As a consequence of Schr\"odinger's equation and of Bohm's law of motion, 
the quantum equilibrium distribution $|\psi(q)|^2$ is equivariant.  This means that if the configuration $Q(t) = (Q_1(t), \ldots, Q_N(t))$ of a system is random with distribution $|\psi_t|^2$ at some time $t$, then this will be true also for any other time $t$. Thus, the \emph{quantum equilibrium hypothesis}, which asserts that whenever a system has wave function $\psi_t$, its configuration $Q(t)$ is random with distribution
\begin{equation}
 \rho = |\psi_t|^2\,,
\end{equation}
can consistently be assumed. This hypothesis is not as hypothetical as its name may suggest: it follows, in fact, by the law of large numbers from the assumption that the (initial) configuration of the universe is typical (i.e., not-too-special) for the $|\Psi|^2$ distribution, with $\Psi$ the (initial) wave function of the universe \cite{DGZ92}. The situation resembles the way Maxwell's distribution for velocities in a classical  
gas follows from the assumption that the phase point of the gas is typical for the uniform distribution on the energy surface.
As a consequence of the quantum equilibrium hypothesis, a Bohmian universe, even if deterministic, appears random to its inhabitants. For a discussion see \cite{DGZ92,DGZ04}.

\subsection{Ghirardi, Rimini, and Weber}\label{sec:GRW}

Ghirardi, Rimini and Weber (GRW) \cite{GRW86} have proposed a nonlinear, stochastic evolution law for quantum mechanical wave functions that deviates from the unitary Schr\"odinger evolution by implementing spontaneous collapses of the wave function. Two primitive ontologies have been proposed for use with the GRW wave function: a \emph{matter density ontology} \cite{Ghi} and a \emph{flash ontology} \cite{Belljumps}, leading to two collapse theories denoted GRWm and GRWf. 

To begin with, the GRW wave function follows a stochastic jump process in Hilbert space.  Consider a quantum system described (in the standard language) by an $N$-``particle'' wave function $\psi = \psi(q_1, \ldots,q_N)$, ${q}_i\in \RRR^3$, $i=1,\dots, N$.  For any point $x$ in $\RRR^3$ (the ``center'' of the collapse that will be defined next), define on the Hilbert space of the system  the \emph{collapse rate operator}
\begin{equation}
\Lambda_i (x) =\frac{1}{(2\pi \sigma^2)^{3/2}}\, e^{-\frac{( \widehat{Q}_i-x)^2}{2\sigma^2}}\,,
\label{eq:collapseoperator}
\end{equation}
where  $\widehat{Q}_i$ is the position operator of  ``particle'' $i$. Here $\sigma$ is a new constant of nature of order of $10^{-7}$m.

Let $\psi_{t_0}$ be the initial wave function, i.e., the normalized wave function at some time $t_0$ arbitrarily chosen as initial time. Then $\psi$ evolves in the following way: 
\begin{enumerate}
\item It evolves unitarily, according to Schr\"odinger's equation, until a random time $T_1= t_0 + \Delta T_1$, so that
\begin{equation}
\psi_{T_1}= U_{\Delta T_1} \psi_{t_0},
\end{equation}
where $U_t$ is the unitary operator $U_t=e^{-\frac{i}{\hbar}Ht}$ corresponding to the standard Hamiltonian $H$ governing the system, e.g., given by (\ref{eq:H}) for $N$ spinless particles, 
and  $\Delta T_1$ is a random time distributed according to the exponential distribution with rate $N\lambda$ (where the quantity $\lambda$ is another constant of nature of the theory,\footnote{Pearle and Squires \cite{PS94} have argued that $\lambda$ should be chosen differently for every ``particle,'' with $\lambda_i$ proportional to the mass $m_i$.} of order of $10^{-15}$ s$^{-1}$).
\item At time $T_1$ it undergoes an instantaneous collapse with random center 
 $X_1$ and random label $I_1$ according to
\begin{equation}
\psi_{T_1} \mapsto\psi_{T_1+}= \frac{\Lambda_{I_1} (X_{1})^{1/2}\psi_{T_1}}{\| \Lambda_{I_1} (X_{1})^{1/2} \psi_{T_1} \|}.
\end{equation}
$I_1$ is chosen at random in the set $\{1, \ldots, N\}$ with uniform distribution. The center  of the collapse 
$X_1$ is chosen randomly with probability distribution
\begin{equation}\label{p}
\mathbb{P}(X_1\in dx_{1}|  \psi_{T_1}, I_1=i_1) = \left\langle \psi_{T_1}|\Lambda_{i_1}(x_1)\psi_{T_1}\right\rangle dx_{1} = \|\Lambda_{i_1} (x_1)^{1/2} \psi_{T_1}\|^2 dx_{1}.
\end{equation}
\item Then the algorithm is iterated: $\psi_{T_1+}$ evolves unitarily until a random time $T_2 =  T_1 + \Delta T_2$, where  $\Delta T_2$ is a random time (independent of $\Delta T_1$) distributed according to the exponential distribution with rate $N\lambda$, and so on.
\end{enumerate}

In other words, the evolution of the wave function is the Schr\"odinger evolution interrupted by collapses. When the wave function is $\psi$, a collapse with center $x$ and label $i$ occurs at rate 
\begin{equation}\label{rate}
r(x,i|\psi)=\lambda\left\langle\psi|\Lambda_{i}(x)\,\psi\right\rangle
\end{equation}
and when this happens, the wave function changes to ${\Lambda_{i} (x)^{1/2}\psi}/{\| \Lambda_{i} (x)^{1/2} \psi \|}$.

In the subsections below I describe GRWm and GRWf, two theories that share the GRW wave function but differ in their postulate about matter. They introduce different kinds of primitive ontology.

\subsubsection{GRWm}
\label{sec:GRWm}

GRWm \cite{Ghi,BG03,Tum05b,AGTZ06} postulates that there is a continuous distribution of matter in space whose density at location $x\in \RRR^3$ and time $t$ is given by
\begin{equation}\label{mdef}
 m(x,t) = \sum_{i=1}^N m_i \int\limits_{\RRR^{3N}}  dq_1 \cdots dq_N \, \delta(q_i-x) \,  \bigl|\psi(q_1, \ldots, q_N,t)\bigr|^2 \,.
\end{equation}
In words, one starts with the $|\psi|^2$--distribution in configuration
space $\RRR^{3N}$, then obtains the marginal distribution of 
the $i$-th degree of freedom $q_i\in \RRR^3$
by integrating out all other variables $q_j$, $j \neq i$, multiplies by the mass associated with $q_i$, and sums over $i$. 
For further discussion of this ontology see \cite{AGTZ06}.

\subsubsection{GRWf}
\label{sec:GRWf}

GRWf was first suggested by Bell \cite{Belljumps,Bellexact}, and then adopted in \cite{Kent89,Tum05a,Mau05,Tum05b,AGTZ06}, for the purpose of obtaining a relativistic collapse theory. According to GRWf, the primitive ontology is given by ``events'' in space-time called flashes, mathematically described by points in space-time. In GRWf matter is neither made of particles following world lines, such as in classical or Bohmian mechanics, nor of a continuous distribution of matter such as in GRWm, but rather of discrete points in space--time, in fact finitely many points in every bounded space-time
region. ``A piece of matter then is a galaxy of such events.''~\cite{Belljumps}

In the GRWf theory, the space-time locations of the flashes can be read
off from the history of the wave function: every flash corresponds to one of the spontaneous collapses of the wave function, and its space-time location is just the space-time location of that collapse. The flashes form the set
\[
  F=\{(X_{1},T_{1}), \ldots, (X_{k},T_{k}), \ldots\}
\]
(with $T_1<T_2<\ldots$).

Note that if the number $N$ of the degrees of freedom in the wave function is large, as in the case of a macroscopic object, the number of flashes is also large (if $\lambda=10^{-15}$ s$^{-1}$ and $N=10^{23}$, we obtain $10^{8}$ flashes per second).
Therefore, for a reasonable choice of the parameters of the GRWf theory, a cubic centimeter of solid matter contains more than $10^8$ flashes per second.
That is to say that large numbers of flashes can form macroscopic shapes, such as tables and chairs. That is how we find an image of our world in GRWf. 
According to GRWf, the wave function serves as the tool to generate
the ``law of evolution'' for the flashes: equation \eqref{rate} gives the rate of the flash process---the probability per unit time of the flash of label $i$ occurring at the point $x$.  
Since the wave function $\psi$ evolves in a random way, $F=\{(X_k,T_k): k\in \NNN\}$ is a random subset of space-time, a point process in space-time. 

Note that GRWm and GRWf, though they share the same wave function and are empirically equivalent, are clearly different theories. For example, one says that matter is continuously distributed, and the other that matter is concentrated in countably many space-time points. As a less trivial example, GRWf allows strong superselection rules, as does Bohmian mechanics, but GRWm does not \cite{CDT05}. As another example, in GRWf but not in GRWm the probability distribution of the history of the primitive ontology is quadratic in $\psi$ \cite{AGTZ06}. As an interesting  consequence, two ensembles of wave functions with the same density matrix have the same distribution of the history of the primitive ontology in GRWf but not in GRWm (neither in Bohmian mechanics). Thus, GRWf is an example of a theory for which one can indeed argue that there is no physical difference between two ensembles of worlds with different distributions of the initial wave function but equal density matrices.

\section{Recent Developments}

Now that the days in which quantum mechanics was mysterious are over, new challenges arise from relativity, from quantum field theory (QFT), and from the combination of the two. And, finally, from quantum gravity. In this section, I report about some recent progress concerning the problems how to extend Bohmian mechanics and the GRW theories to quantum field theory and how to make them relativistic. I first turn to quantum field theory, and then to the relativity question.

\subsection{Bohmian Mechanics and Quantum Field Theory}

Two ways of extending Bohmian mechanics to quantum field theory are known: either by postulating that a \emph{field configuration} (instead of a \emph{particle configuration}) is guided by a wave function (understood as a functional on the field configuration space) \cite{Bohm52,BH93}, or by introducing particle creation and annihilation into Bohmian mechanics \cite{Bell86,crlet,crea2B}. The latter approach is called ``Bell-type quantum field theory'' since the first model of this kind (on a lattice) was proposed by Bell \cite{Bell86}.

\subsubsection{Field Ontology}

Although the field ontology was proposed already in the 1950s \cite{Bohm52} and taken up by several authors \cite{Val,BH93,Hol95,HD04,Str05,SW06}, it has not been sufficiently developed to clarify whether it provides a viable theory. There exist no rigorous studies of this approach, and in particular the obvious question how to obtain an equivariant measure, the analog of $|\psi(q)|^2 dq$, on an infinite-dimensional configuration space where no Lebesgue volume measure exists, has not been addressed. (This question is not of the same importance in orthodox quantum field theory, as normal experiments concern the detection of particles rather than the measurement of the field at all points of space.)

Another question that needs to be addressed is whether what we normally regard as different macrostates is actually supported by disjoint field configurations. This property is not obvious if the primitive ontology is not directly related to (what can be regarded as) the distribution of matter in space (for example, for an ontology of electromagnetic fields \cite{SW06}). And this property is relevant for ensuring that measurement results get displayed and recorded in the primitive ontology.

\subsubsection{Bell-Type Quantum Field Theories}
\label{sec:BTQFT}

In Bell-type QFTs, the motion of the configuration along deterministic trajectories is interrupted by stochastic jumps, usually corresponding to the creation or annihilation of particles. A typical example of a configuration space in this context is the configuration space $\Q=\Gamma(\RRR^3)$ of a variable number of identical particles, which can be defined as the set of all finite subsets of 3-space, or, equivalently, as the (disjoint) union of all $N$-particle configuration spaces
\begin{equation}\label{NRtredef}
  {}^{N}\RRR^3 := \{q \subset \RRR^3: \#q = N\} = \bigl(\RRR^{3N} 
  \setminus \{\text{coincidences}\} \bigr)/\text{permutations}\,.
\end{equation}
A history of particles in $\RRR^3$ that can be created, move, and be annihilated corresponds to a path $t \mapsto Q_t$ in $\Q$ that jumps, at every time of creation or annihilation, from one sector ${}^N\RRR^3$ to another. 

In a Bell-type QFT, the configuration $Q_t$ follows a Markov jump process in $\Q$. This means, in every time interval $[t,t+dt]$ the configuration $Q_t$ has probability
\begin{equation}
  \sigma_t(Q_t\to q) \, dt \, dq
\end{equation}
to jump to the volume $dq$ around the configuration $q$, and in case it does not jump it moves continuously according to Bohm's law of motion
\begin{equation}
  \frac{dQ_t}{dt} = v^{\psi_t} (Q_t) \,.
\end{equation}
The \emph{jump rate} $\sigma(q' \to q)\, dq$ (probability per time) is prescribed by the following law  in terms of the wave function $\psi$, which is usually from Fock space:
\begin{equation}\label{Bell}
  \sigma^\psi(q' \to q) = \tfrac{2}{\hbar} 
  \frac{[\Im \,\langle \psi | q \rangle \langle q | H_I | q' \rangle 
  \langle q' | \psi \rangle]^+}{\langle \psi | q' \rangle \langle q' | \psi \rangle} \,,
\end{equation}
where $H_I$ is the interaction Hamiltonian and $s^+ = \max(s,0)$ denotes the positive part of $s \in \RRR$. 

This law is dictated by the following considerations: We want the measure $|\psi(q)|^2 dq$ to be equivariant, i.e., we want that $Q_t$ has distribution $|\psi_t|^2$ provided $Q_0$ had distribution $|\psi_0|^2$. Moreover, in rough analogy to the formula 
\begin{equation}
 j = \tfrac{\hbar}{m} \Im (\psi^* \nabla \psi)
\end{equation}
for the probability current in quantum mechanics there is a formula
\begin{equation}
  J(q,q') = \tfrac{2}{\hbar} \Im \,\langle \psi | q \rangle \langle q | H_I | q' \rangle 
  \langle q' | \psi \rangle
\end{equation}
for the probability current between $q$ and $q'$ due to $H_I$, and we want $J(q,q') \, dq\, dq'$ to be the amount of probability flowing (per time) from $dq'$ to $dq$ minus the amount from $dq$ to $dq'$. Among all jump rates $\sigma$ with this property, \eqref{Bell} is the smallest, leading to no more jumps than necessary for ensuring the prescribed net flow of probability. This is the reason why the jump process with rate \eqref{Bell} is called the \emph{minimal process} associated with $\psi$ and $H_I$. 

It has been made plausible \cite{crea2B} that this concept extends to a natural way of associating, with every Hamiltonian $H$ (from a large class of operators) and every initial wave function $\psi$ (from a dense subspace) evolving according to $H$, a Markov process in configuration space, the \emph{minimal process}. More precisely, it is the triple
\begin{equation}
  \Bigl(\text{Hilbert space, Hamiltonian, position operators} \Bigr) 
\end{equation}
that defines, for every $\psi$, a process $Q^\psi$. The ``position operators'' are given as a positive-operator-valued measure (POVM) on configuration space. Thus, whenever a QFT  is given as such a triple, there is a Bohm-like process in $\Q$ associated with it in a canonical way. Of course, this scheme does not provide particle paths for QFTs with ill-defined Hamiltonians. Here are some examples of minimal processes:
\begin{itemize}
\item The minimal process associated with the Schr\"odinger operator \eqref{eq:H} is Bohmian mechanics.
\item The one associated with the Dirac operator
\begin{equation}
  H=-ic\hbar {\alpha} \cdot \nabla + mc^2 \beta
\end{equation}
is Bohm's 1953 \cite{Bohm53} law of motion for a Dirac particle,
\begin{equation}\label{BohmDirac}
  \frac{dQ}{dt} = \frac{\psi^\dagger \alpha \psi}{\psi^\dagger \psi}(Q) \quad \text{or} \quad
  \frac{dX^\mu}{d\tau} \propto j^\mu(X(\tau))\,,\quad 
  j^\mu = \overline{\psi} \gamma^\mu \psi\,,
\end{equation}
where $Q(t) \in \RRR^3$ is the position of the particle at time $t$, $X^\mu =(t,Q(t))$ is the corresponding space-time point, and $\tau$ an arbitrary curve parameter (e.g., proper time).
\item The minimal process associated with integral operators $H=H_I$ is the pure jump process with rates \eqref{Bell}.
\item The one associated with $H=H_0 + H_I$, the sum of a Schr\"odinger or Dirac operator $H_0$ and an integral operator $H_I$, is Bohmian motion interrupted by stochastic jumps.
\item For quantum mechanics on a graph, a candidate process is known \cite{graph}.
\end{itemize}

Here is a look at the literature. The jump rate \eqref{Bell} was first considered by Bell \cite{Bell86} on the lattice and in \cite{crea1} in the continuum, and further discussed in \cite{Sudbery,Vink,Den03,crlet,crea2A,crea2B,Col04,Col05, GT05a,GT05b}. Some mathematical works study the following aspects: conditions under which the jump rate is rigorously defined \cite{crea2A}, conditions for the global existence of Bell's lattice process \cite{GT05a}, conditions for the global existence of the trajectories in the absence of jumps \cite{TT05}; a global existence proof for the combination of continuous motion and jumps is still missing but seems doable. It has been conjectured \cite{Sudbery,Vink,crea2B} (but not rigorously proven) that Bell's process for the lattice approximation to the Schr\"odinger equation converges to Bohmian mechanics as the lattice width goes to zero. It has been observed in \cite{CDT05} that in Bell-type QFTs, superselection rules sometimes hold in the strong sense that every superposition (relative to the superselected observable) can be replaced by a mixture without changing the probability distribution on path space, as opposed to the weak sense of superselection in which every superposition can be replaced by a mixture without an empirically detectable difference. For a proposal how to make Bell's lattice process deterministic by introducing further variables see \cite{HCD04}.

\subsubsection{Position Operators}\label{sec:POVM}

For the choice of configuration space and the position operators, there is often an obvious candidate. When we deal with several species of particles, we may take the configuration space $\Q$ to be the Cartesian product of several copies of $\Gamma(\RRR^3)$, and the position POVM to be the corresponding product \cite{crea2B}. For the quantized Dirac field, an obvious candidate, used by D\"urr \textit{et al.}\ \cite{crea2B}, would be $\Q = \Gamma(\RRR^3) \times \Gamma(\RRR^3)$, so that a configuration specifies some electron points and some positron points, with the position POVM determined on each factor by the electron (positron) number operators on the appropriate Fock spaces, often denoted $b^\dagger(\vr) \, b(\vr)$ and $d^\dagger(\vr) \, d(\vr)$ with $\vr \in \RRR^3$. 

A different proposal for the quantized Dirac field, corresponding primarily to a different choice of position operators, has been made by Bohm and Hiley \cite[p.~276]{BH93} and, more recently and in more detail, by Colin \cite{Col04,Col05}, so I will call it ``Colin's picture'' in the following. He proposes to take the ``Dirac sea'' literally and to introduce infinitely many particles, each having a trajectory, obeying the analogue of \eqref{BohmDirac} with a wave function of infinitely many particles. (This wave function should be ``the filled Dirac sea'' with finitely many electrons removed from the negative energy states and/or added with positive energy states.) This kind of Bohmian dynamics for infinitely many particles has been defined only heuristically, without mathematical rigor; since the configuration space becomes infinite-dimensional and such spaces do not possess a Lebesgue volume measure, it remains unclear whether and how an equivariant measure, the analogue of $|\psi(q)|^2\, dq$, can be defined. Since in Colin's picture, every little volume of space contains infinitely many particles, it is not obvious how to read off pointer positions, or, more generally, how to obtain from it a familiar picture of matter, in which (say) tables and chairs are discernible. To this end, Colin \cite{Col05} has proposed a certain way of coarse-graining the density of matter, and has argued heuristically that for a typical configuration of a state consisting of the filled Dirac sea with one electron added, the coarse-grained density has a discernible peak over the ``sea level,'' a peak of height 1 and width the Compton wave length.

Three differences between Colin's picture ($C$) and that of D\"urr \textit{et al.}\ ($D$) can be mentioned. First, $C$ is deterministic, and $D$ is stochastic. Second, a drop of water in $D$ consists of about $10^{24}$ particles, which agrees with what one would naively expect before worrying about quantum field theory. In $C$, it consists of infinitely many particles belonging to the Dirac sea, as does the same volume of vacuum. This trait of $C$ is not inacceptable but a bit eccentric. Third, in usual quantum field theory there is a symmetry between electrons and positrons, in the sense that one could just as well regard the electron as the anti-particle of the positron. This symmetry is respected in $D$ and broken in $C$, in which electrons are real but positrons are merely holes.

Let me turn to another aspect of the choice of the position POVM. Goldstein \textit{et al.}\ \cite{aapi} have elaborated on the possibility first considered by Bell \cite{Bell86} that ``particles are just points,'' which means that the primitive ontology does not include intrinsic differences between particles of different species, and entails that, even for particles of different species, the configuration space is that of identical particles, $\Gamma(\RRR^3)$. This can always be arranged by suitably projecting the position POVM to $\Gamma(\RRR^3)$. Alternatively, this POVM arises in a canonical way from the particle number operators $N(\vr)$ \cite[Sec.~6.8]{crea2B}.

Despite the differences between the various pictures, including the field ontology, it is striking that they have in common: (i)~the attitude about what a theory has to achieve for being acceptably clear; (ii)~the structure of matter being described by beables and guided (if only stochastically) by the wave function; (iii)~the status of the observables as being secondary to the beables; and (iv)~the unitary evolution of the wave function.

\subsection{GRW and Quantum Field Theory}

Since there are no particles in the GRW theories, configuration space plays not the same role as in Bohmian mechanics, and a POVM on configuration space is not the relevant mathematical object. Instead, for the purposes of GRW theories, a QFT can be thought of as given by the triple
\begin{equation}
  \Bigl( \text{Hilbert space, Hamiltonian, 
  matter density operators }\mathcal{M}(\vr) \Bigr) \,.
\end{equation}
For the matter density one could take either the particle number density operators $\mathcal{M}(\vr) = N(\vr)$, such as $\mathcal{M}(\vr) = b^\dagger(\vr) \, b(\vr) + d^\dagger(\vr) \, d(\vr)$ for the quantized Dirac field, or (preferably, say \cite{PS94,BG03}) the mass density operators. Then the collapse rate operators $\Lambda(\vr)$ are obtained by convolution with a Gaussian of width $\sigma$,
\begin{equation}
  \Lambda(\vr) = \int d^3 \vr' \, \mathcal{M}(\vr') \, \frac{1}{(2\pi \sigma^2)^{3/2}}
  \, e^{-\frac{( \vr-\vr')^2}{2\sigma^2}}\,.
\end{equation}
Given these operators, there is a canonical GRW-like collapse process with collapse rate \eqref{rate} \cite{Tum05b, Tum06}, and a canonical CSL process \cite{GPR90} (continuous spontaneous collapse; see \cite{BG03} for a review). The GRW-like process can be combined with either the flash ontology or the matter density ontology, the CSL process with the matter density ontology.

\subsection{Bohmian Mechanics and Relativity}\label{sec:relBohm}
\subsubsection{The Time Foliation}\label{sec:foliation}

With the invocation of a preferred foliation $\foliation$ of space-time into spacelike 3-surfaces, given by a Lorentz invariant law, it is known \cite{HBD} that Bohmian mechanics possesses a natural generalization to relativistic space-time. I will call this foliation the \emph{time foliation} in the following, to distinguish it from all the other foliations that a Lorentzian manifold possesses, and the 3-surfaces belonging to $\foliation$ the \emph{time leaves}. The role of the time foliation is to define which configurations of $N$ space-time points should be counted as ``simultaneous'' when plugging the ``simultaneous'' positions of all particles into (the analog of) Bohm's equation of motion. The possibility of a preferred foliation seems against the spirit of relativity (see \cite{timbook} for a discussion), but certainly worth exploring. It is suggested by the empirical fact of quantum non-locality and by the structure of the Bohmian law of motion \eqref{Bohm} for many particles, in which the velocity of a particle depends on the simultaneous positions of the other particles. The GRW theory differs from Bohmian mechanics in that it can be made relativistic without the invocation of a time foliation (see Section~\ref{sec:relGRW} below). 

Using a time foliation, a Bohm-type equation of motion was formulated by D\"urr \textit{et al.}\ \cite{HBD} for flat space-time (based on earlier work in \cite{BH93,DGZ90}; for a lattice version see \cite{samols}; for a version with a field ontology see \cite{HD04}; the straightforward generalization to curved space-time was formulated and mathematically studied in \cite{3forms}):
\begin{equation}\label{hbd}
  \frac{dX_k^{\mu_k}}{d\tau} \propto j^{\mu_1 \ldots \mu_N} 
  \bigl(X_1(\Sigma),\ldots, X_N(\Sigma)\bigr) 
  \prod_{i\neq k} n_{\mu_i}\bigl(X_i(\Sigma)\bigr)\,,
\end{equation}
where $X_k(\tau)$ is the world line of particle $k\in \{1,\ldots,N\}$, $\tau$ is any curve parameter, $\Sigma$ is the time leaf containing $X_k(\tau)$, $n(x)$ is the unit normal vector on $\Sigma$ at $x \in \Sigma$, $X_i(\Sigma)$ is the point where the world line of particle $i$ crosses $\Sigma$, and
\begin{equation}\label{multij}
  j^{\mu_1 \ldots \mu_N} = \overline{\psi} (\gamma^{\mu_1} \otimes \cdots \otimes
  \gamma^{\mu_N}) \psi
\end{equation}
is the probability current associated with the Dirac wave function $\psi$ defined on $\bigcup_{\Sigma \in \foliation} \Sigma^N$.

As mentioned before, the time foliation might itself be dynamical. It is to be regarded as a physical object, just as the space-time metric or the wave function, and as such should be governed by an evolution law. An example of a possible Lorentz invariant evolution law for the foliation is
\begin{equation}\label{lawF}
  \nabla_\mu n_\nu - \nabla_\nu n_\mu =0\,.
\end{equation}
This law allows to choose an initial spacelike 3-surface and then determines the foliation. It is equivalent to saying that the infinitesimal timelike distance between two nearby 3-surfaces from the foliation is constant along the 3-surface. As a consequence, there is a system of space-time coordinates $x^0, \ldots, x^3$ such that $x^0$ is constant on every time leaf, and
\begin{equation}
  g_{\mu\nu} = \left( \begin{matrix} 1& 0 \\ 0 & g_{ij}  \end{matrix} \right)
\end{equation} 
with $g_{ij}$ a Riemannian 3-metric.

A special foliation $\foliation_\mathrm{BB}$ obeying \eqref{lawF} is the one consisting of the surfaces of constant timelike distance from the Big Bang (i.e., the initial singularity). It is the foliation defined by the function that could be regarded as the only natural concept of ``absolute time'' available in a Lorentzian manifold with Big Bang; for example, the ``absolute time'' of here-now is $13.7 \pm 0.2$ billion years \cite{wmap}.

Actually, the law of motion \eqref{hbd} does not require any particular choice of law for the foliation, except that the foliation does not depend on the particle configuration (while it may depend on the wave function).

\subsubsection{Other Proposals for Relativistic Bohmian Mechanics}
\label{sec:multit}

Since the existence of a time foliation would be against the spirit of relativity, several attempts have been undertaken at obtaining a relativistic Bohm-like theory without a time foliation. I briefly describe four such proposals in this subsection, items (i)--(iv) below. However, (i)--(iii) are not satisfactory theories, and (i) and (iv) both involve some foliation-like structure, something just as much against the spirit of relativity as a time foliation.

First we need the concept of a \emph{multi-time wave function} $\psi(q_1,t_1, \ldots, q_N,t_N)$, which is the obvious generalization of an $N$-particle wave function $\psi(q_1,\ldots,q_N,t)$ to the relativistic setting. It involves one time variable for every particle and thus constitutes a function on $(\text{space-time})^N$. For $N$ time variables one needs $N$ Schr\"odinger equations
\begin{equation}\label{multit}
  i\hbar \frac{\partial \psi}{\partial t_k} = H_k \psi\,,
\end{equation}
and such a set of equations cannot always consistently be solved.\footnote{To begin to understand why, consider for simplicity $N=2$, suppose you specify initial data for $t_1= t_2=0$ and think of the multi-time wave function as a Hilbert-space-valued function of $(t_1,t_2)$, with the Hilbert space $L^2(q_1, q_2)$. Then note that, in order for such a function of $(t_1,t_2)$ to exist, evolving the vector in Hilbert space from time $(0,0)$ to (say) $(t,0)$ and then from $(t,0)$ to $(t,t)$ must lead to the same result as first evolving from $(0,0)$ to $(0,t)$ and then from $(0,t)$ to $(t,t)$.} The condition for consistency reads:
\begin{equation}\label{multitconsistency}
  \biggl[\frac{i\hbar\partial}{\partial t_k} - H_k, 
  \frac{i\hbar\partial}{\partial t_j} - H_j \biggr] = 0 \quad \text{for } k \neq j\,.
\end{equation}
In quantum mechanics, this condition is satisfied for non-interacting particles but not in the presence of an interaction potential. It seems that consistent multi-time equations with interaction are possible if the interaction is implemented, not by a potential, but by creation and annihilation of other particles. 

In the remainder of this section I consider only non-interacting particles, and the corresponding unitary evolution is given by $N$ Dirac equations,
\begin{equation}\label{multitDirac}
  i\hbar \gamma^\mu_k \frac{\partial \psi}{\partial x_{k}^{\mu}} 
  = m_k \psi\,,
\end{equation}
where $m_k$ is the mass of the $k$-th particle, and $\gamma^\mu_k$ is the Dirac gamma matrix $\gamma^\mu$ acting on the spin index of the $k$-th particle. Now I turn to the four proposed relativistic modifications of Bohmian mechanics.

\begin{itemize}
\item[(i)] \textit{Synchronized trajectories} \cite{BDGZ96,DH01,nikolic}. Define a path $s \mapsto X(s)$ in $(\text{space-time})^N$ as the integral curve of a vector field $j^\psi$ on $(\text{space-time})^N$, with $j^\psi$ a suitably defined current vector field obtained from a wave function $\psi$ on $(\text{space-time})^N$. The path $X(s) = \bigl(X_1(s), \ldots, X_N(s) \bigr)$ defines $N$ paths in space-time, parametrized by a joint parameter $s$, which are supposed to be the particle world lines. This approach is based on a naive replacement of space with space-time. Apparently, it does not possess any equivariant measure, and thus does not predict any probabilities. Moreover, it does introduce a foliation-like structure: The joint parametrization defines a \emph{synchronization} between different world lines, as it defines which point on one world line is simultaneous to a given (spacelike separated) point on a second world line. Indeed, the synchronization is encoded in the world lines since, if $N$ non-synchronous points $X_1(s_1), \ldots, X_N(s_N)$ on the $N$ world lines are chosen, then the integral curve $s\mapsto Y(s)$ of $j^\psi$ starting from $Y(0) = \bigl( X_1(s_1), \ldots, X_N(s_N) \bigr)$ will generically lead to different world lines than $X$.

\item[(ii)] \textit{Light cones as simultaneity surfaces} \cite{GT03}. Using as surfaces of simultaneity the future or past light cones, one can specify a Bohm-like equation of motion for $N$ particles without a time foliation or similar structure. If using past light cones, the theory is local, but if using future light cones it is nonlocal, thus providing a toy example of a relativistic nonlocal theory. Since its equation of motion is a sort of differential delay equation with advanced arguments, it possesses a microscopic arrow of time pointing towards the past, i.e., opposite to the macroscopic (thermodynamic) arrow of time. Apparently, this theory does not possess any equivariant measure, and thus does not predict any probabilities.

\item[(iii)] \textit{Covariant velocity vector fields} \cite{GT01,first}. Consider $N$ particles, with positions $(Q_1(t), \ldots,Q_N(t)) =: Q(t)$, moving according to the equation of motion $dQ/dt = v(Q)$, with $v$ a vector field on $\RRR^{3N}$. Suppose $v$ has the property that every integral curve $t\mapsto Q(t)$, when understood as $N$ curves in space-time, will transform under any Poincar\'e transformation into another integral curve of $v$. Then the theory is covariant, without invoking a time foliation or similar structure. I can show that such ``covariant'' vector fields $v$ exist for every $N\geq 3$, and that the resulting particle theory is nonlocal. In such a theory, any Lorentz frame could equally be regarded as providing the surfaces of simultaneity used in the law of motion. However, it is not clear how to obtain any probabilities from such theories, as they do not provide a measure on the space of solution curves $t \mapsto Q(t)$. In addition, they have a kind of conspiratorial character, as a consequence of which they are very incompatible with free will, more so than other deterministic or stochastic theories. (For example, the theoretical treatment of a system of nonrelativistic Bohmian particles allows external potentials to be treated as free variables, at the whim of the experimenter, as long as the experimenter herself is not included in the deterministic treatment.)

\item[(iv)] \textit{Flashes with the Schr\"odinger evolution.} This model, described in \cite{AGTZ06} for a different purpose in a nonrelativistic setting under the name $Sf'$, uses the flash ontology, but (unlike GRWf) is empirically equivalent to orthodox quantum mechanics. Consider a relativistic system of $N$ noninteracting quantum particles with multi-time wave function governed by $N$ Dirac equations \eqref{multitDirac}. Each of the flashes is associated with one of the particle labels $1, \ldots, N$, and one ``seed flash'' $X_k$ of every label must be specified as part of the initial data, together with a (timelike, future-pointing) unit vector $u_k^\mu$ from the tangent space at $X_k$. Then one can devise a covariant algorithm for constructing the subsequent random flashes, each with a unit tangent vector, by plugging the previous flashes into the other variables of the wave function.\footnote{Here is how. Choose a random value $T_1$ with exponential distribution with expectation $1/\lambda$ and a random space-time point $\tilde X_1$, the next flash with label 1, on the 3-surface $\Sigma_1$ of constant timelike distance $T_1$ from $X_1$ with probability distribution 
\begin{equation}
  \PPP(\tilde X_1 \in d^3x_1) = \mathcal{N}_1 \, 
  j^{\mu_1 \ldots \mu_N} (x_1,X_2, \ldots, X_N) \, 
  n_{1,\mu_1}(x_1) \, u_{2,\mu_2} \cdots u_{N,\mu_N}
  \, \mathrm{Vol}(d^3x_1)\,,
\end{equation}
where $j^{\mu_1 \ldots \mu_N}$ is defined by \eqref{multij}, $\mathrm{Vol}(d^3x_1)$ is the Riemannian 3-volume measure on the surface $\Sigma_1$, 
$n_{1,\mu}$ the unit normal vector field on $\Sigma_1$, 
and $\mathcal{N}_1$ a normalizing constant. Set $\tilde u_{1,\mu} = n_{1,\mu}(\tilde X_1)$. Repeat the same procedure with label 2, i.e., choose a random value $T_2$ and a random point $\tilde X_2$ on the 3-surface $\Sigma_2$ of timelike distance $T_2$ from $X_2$ with distribution 
\begin{multline}
  \PPP(\tilde X_2 \in d^3x_2|\tilde X_1, \tilde u_1) = \mathcal{N}_2 \, 
  j^{\mu_1 \ldots \mu_N} (\tilde X_1, x_2, X_3, \ldots, X_N) \:\times \\ 
  \tilde u_{1,\mu_1} \, n_{2,\mu_2}(x_2) \, u_{3,\mu_3} \cdots u_{N,\mu_N} 
  \, \mathrm{Vol}(d^3x_2)\,,
\end{multline}
and so on. (Alternatively, choose at random which label to proceed with.)}
This theory arguably reproduces the quantum mechanical probabilities, or at least it would if interaction were incorporated. A trait of this theory that is absent from relativistic GRWf is that the flashes are endowed with a temporal ordering, defining which of two flashes at spacelike distance is earlier and which is later. This is because the flashes are constructed here in \emph{generations}, and the distribution of a flash depends upon which of the other flashes belong to the same or the previous generation. Thus, this theory also contains some foliation-like structure, but at least it works better than the theory with synchronized trajectories, as it yields the right probabilities.

\end{itemize}

\subsection{GRW and Relativity}
\label{sec:relGRW}

The GRW theory can be made relativistic, without a time foliation or any similar structure, when using the flash ontology \cite{Tum05a}. This was conjectured first by Bell \cite{Belljumps}; for discussions of the relativistic GRWf model see also \cite{Mau05,Tum06,AGTZ06}. A relativistic collapse model on a lattice has been described by Dowker and Henson \cite{dowker1}.

At present, the relativistic GRWf model is known only for $N$ non-interacting ``particles'' as it uses multi-time wave functions (as discussed in Section \ref{sec:multit}). The defining equations of the relativistic GRWf model are spelled out in \cite{Tum05a,Tum06}; here I limit myself to describing its structure. Each flash has a label or ``type'' $i\in\{1,\ldots,N\}$. Choose an arbitrary spacelike 3-surface $\Sigma_0$ where initial conditions are specified. As the initial conditions, specify a (normalized Dirac) wave function $\psi_{\Sigma_0}$ on $\Sigma_0^N$ and one ``seed flash'' of every type in the past of $\Sigma_0$, to be thought of as the last flash of its type before $\Sigma_0$. Then the model specifies, by its defining law, the joint probability distribution of all flashes (and their types) in the future of $\Sigma_0$. This law is independent of the choice of coordinates and does not require or generate a time foliation or any similar structure.

The foliation independence of the model can be expressed in the following way: With every spacelike 3-surface $\Sigma$ in the future of $\Sigma_0$ there is associated a wave function $\psi_\Sigma$ on $\Sigma^N$, the \emph{conditional wave function}, which depends on all flashes between $\Sigma_0$ and $\Sigma$, as well as on the seed flashes before $\Sigma_0$ and, of course, on the initial wave function. (Indeed, the conditional wave function collapses at every flash.) Then the conditional probability distribution of all flashes (and their types) in the future of $\Sigma$, given the flashes between $\Sigma_0$ and $\Sigma$, coincides with the distribution given by the model's defining law, with the initial 3-surface $\Sigma_0$ replaced by $\Sigma$, $\psi_{\Sigma_0}$ replaced by $\psi_\Sigma$, and the seed flashes replaced by the last flash of every type before $\Sigma$.\footnote{The equation to be used here as the defining law is (33) in \cite{Tum05a}, or (29) in \cite{Tum06}. One can obtain simpler formulas, equation (19) in \cite{Tum05a} or (21) in \cite{Tum06}, for defining the joint distribution of all flashes, if either the seed flashes lie on $\Sigma_0$, or if one dispenses with any initial 3-surface $\Sigma_0$ by specifying (in addition to the seed flashes) the pre-collapse wave function not on $\Sigma_0$ but on all of $(\text{space-time})^N$, i.e., by specifying on (space-time)$^N$ what the wave function would have been if no flash after the seed flashes had ever occurred.}

For understanding the model it is important to realize that the matter (or primitive ontology) is given by the flashes, whereas the wave function has a different status: that of a physical object influencing the matter. Suppose $\Sigma$ and $\Sigma'$ are two different spacelike 3-surfaces having a large portion $\Sigma \cap \Sigma'$ in common, then $\psi_\Sigma$ and $\psi_{\Sigma'}$ can be quite different, due to collapses at flashes between $\Sigma$ and $\Sigma'$. For example, an EPR--Bell pair could be in a singlet state on $\Sigma$ but have collapsed to a product state on $\Sigma'$. In particular, even the reduced density matrices pertaining to the region $\Sigma \cap \Sigma'$ (obtained from $\psi_\Sigma$ respectively $\psi_{\Sigma'}$ by a partial trace) could be different, such as, in the example, ($\tfrac{1}{2}$ times) a two-dimensional respectively a one-dimensional projection. Had we left the primitive ontology unspecified, or regarded the wave function as the primitive ontology, it would have appeared profoundly problematical that the theory does not associate a unique quantum state with a piece of 3-surface such as $\Sigma \cap \Sigma'$. But no problem arises from this fact in GRWf because the behavior of matter, constituted by the flashes, is always unambiguous. For the same reason, no conflict with relativity arises from the fact that in every coordinate system $(x^0, \ldots, x^3)$ on space-time, the collapse of $\psi_t =  \psi_{\{x^0 =t\}}$ takes place instantaneously over arbitrary distances, along any level surface $\{x^0 = t\}$ containing a flash.

Observe also that the same wave function with a different primitive ontology, the matter density ontology, would not be relativistic, at least not with a naive application of \eqref{mdef}. Thus, one cannot decide whether a collapse model is relativistic or not until the primitive ontology is clearly specified. I see this as the main obstacle that previous attempts at defining relativistic collapse theories encountered. 

For example, also the CSL approach succeeds (if we leave aside problems with divergences) in attributing, in a Lorentz invariant way without a time foliation or similar structure, to every spacelike 3-surface $\Sigma$ a collapsed wave function $\psi_\Sigma$. As with relativistic GRWf, the reduced states on $\Sigma \cap \Sigma'$ obtained from $\psi_\Sigma$ and $\psi_{\Sigma'}$ may differ. In addition to the evolution law for $\psi_{\Sigma}$, Ghirardi \cite{Ghi99} has mentioned what he calls a ``criterion for events'' but has not made its status completely clear: Should we regard it as a consequence arising from an analysis of $\psi_\Sigma$ (as we might in GRWf) or as a postulate introducing a primitive ontology $\xi$ and a law for $\xi$ (i.e., as an alternative to GRWf)? In any case, the ``criterion'' asserts that if $A$ is a local observable associated with space-time point $x$ and $PLC(x)$ is the past light cone of $x$, then $A$ is attributed the value $\alpha$ if
\begin{equation}
  A\psi_{PLC(x)} = \alpha \psi_{PLC(x)}\,, 
\end{equation}
otherwise $A$ is attributed the value ``indefinite.'' (Although $PLC(x)$ is not spacelike, it is a limit of spacelike 3-surfaces, so we may hope everything is well defined.) In order to have a clear primitive ontology, we may take $\xi$ to be these ``values'' for all local observables $A$. With this attitude it becomes clear why Ghirardi \cite{Ghi99} so strongly rejected criticisms on the grounds that different spacelike surfaces $\Sigma,\Sigma'$ sometimes attribute different reduced states to $\Sigma \cap \Sigma'$. After all, the criticism refuses to pay attention to a crucial part of the ontology, namely $\xi$.

However, we then realize that once we have postulated that, like in GRWm, the density of matter is given in this way with, say, $A=\mathcal{M}(x)$ the mass density at $x$, all other ``values'' are of no relevance. They are superfluous, as they do not influence how much matter is where, and thus do not influence the positions of pointers or the shape of ink on paper. They are truly \emph{hidden} variables and, indeed, can be deleted from the theory without unpleasant consequences, just like the ether in relativistic mechanics, and unlike the particles in Bohmian mechanics.
Thus, if we take the primitive ontology seriously, we should restrict the ``criterion for events'' to, say, the mass density $\mathcal{M}(x)$. Moreover, since we want to regard the ``value'' of $\mathcal{M}(x)$ as the density of matter, we do not want it to be often ``indefinite.'' We may thus prefer to take $\xi$ to be, instead of the eigenvalue of $\mathcal{M}(x)$, the value that would in orthodox quantum mechanics be regarded as its average,
\begin{equation}
  \xi = m(x) = \langle \psi_{PLC(x)} | \mathcal{M}(x) | \psi_{PLC(x)} \rangle\,.
\end{equation}
This theory, if it can be made rigorous, could be regarded as a relativistic version of GRWm.

\section{Outlook: A Bohm-Like Model For Electromagnetism and Gravity}

In this last section, I propose some quite concrete but speculative model for how to include gravity into a Bell-type quantum field theory. Previous proposals for Bohm-like theories of gravity are based on a wave function on ``superspace'' (i.e., the space of all Riemannian 3-geometries), guiding a point $g(t)$ in superspace as the actual geometry of space at time $t$ \cite{Hol95,shtanov,GT01}, but I follow here a different path.

The standard way of obtaining a quantum theory (such as quantum mechanics, quantum electrodynamics (QED), and quantum gravity) is by \emph{quantization} of a known classical theory. I will describe an alternative path, inspired by Bohmian mechanics. It is obvious that quantization as a method of obtaining quantum theories has its limitations, as one would not have guessed the existence of spin, or the Dirac equation, in this way. Even less meaning is attributed to quantization rules from the Bohmian perspective, since the observables are no longer the central objects of the theory (they need not even be mentioned for defining the theory), they are not obtained by a quantization postulate (but emerge from the law of motion as the mathematical objects encoding the statistics of results of experiments), and their non-commutativity is not regarded as the central innovation of quantum theory (because the non-commuting operators are not regarded as a kind of paradoxical reality). From the Bohmian perspective, quantization is rather the inverse operation to taking the classical limit. 

The questions one naturally asks when trying to define a Bohmian theory involve how to write evolution laws for the particles and the wave function guiding them. Thus, for example, the program of finding all covariant linear wave equations (associated particularly with the names of Dirac and Wigner) is more in the Bohmian spirit than quantization.

\subsection{Photons}

In the Bohmian framework, it seems a natural assumption to me that the word ``photon'' refers to an actual particle (with a position, of course!). I recognize that there are other possibilities, such as a field ontology \cite{Bohm52}, or perhaps no beables at all associated with the electromagnetic quantum field \cite{Bell86}. But the most naive, most obvious, and simplest possibility seems that of photon trajectories, given the striking parallels between the behavior of light and that of matter, such as interference and entanglement. Indeed, if I should list the crucial differences between photons and electrons, I would mention mass, charge, and spin; and the mere difference in these parameters does not suggest to me a difference in ontology, such as electrons being particles and photons being fields \cite{Bohm52} (or nothing at all \cite{Bell86,aapr}, or even photons being fields and electrons being nothing \cite{SW06}). I would also include in the list that photons are bosons while electrons are fermions; and again, the mere fact that the wave function is symmetric in one case and anti-symmetric in the other does not suggest to me a difference in ontology. Thus, photon trajectories seem a good starting hypothesis. And indeed, it is not difficult to write down equations for Bohmian trajectories for bosonic mass-0, charge-0, spin-1 particles \cite{GMGS01,NT06}.

\subsection{Dynamical Configuration Space}

Because of the similarity between electromagnetism and gravity, it also seems a good starting hypothesis that there should be graviton trajectories as well. And again, it is not difficult to write down equations for Bohmian trajectories for bosonic mass-0, charge-0, spin-2 particles, as gravitons are supposed to be. But introducing photon and graviton particles is not enough for obtaining a Bohm-like theory of electromagnetism and gravity, for several reasons:
\begin{itemize}
\item[(i)] Such equations assume a \emph{metric} (i.e., a space-time geometry) as given, and we do not want to assume a fixed a \emph{background metric}. Instead, we want a theory of gravity to create its own metric, a \emph{dynamical metric}. The metric is involved, for example, in the connection (i.e., Christoffel symbols) needed for defining the (covariant) derivatives of the wave function that occur in the appropriate Schr\"odinger equation (the Dirac equation for electrons, and others for photons and gravitons).
\item[(ii)] Something similar can be said of electromagnetism, since the derivative that occurs in the Dirac equation involves (a $U(1)$ gauge connection corresponding to) the electromagnetic vector potential. 
\item[(iii)] There is another place where a space-time metric is needed. When we consider Bohmian trajectories in Euclidean 3-space, then the Euclidean geometry is one of the mathematical objects needed for making physical sense of the trajectories. That is why we should count the space-time geometry as part of the primitive ontology. If I told you merely the coordinates of the particles in my favorite coordinate system but not the metric in this coordinate system, you would not know anything useful because there exist diffeomorphisms $\RRR^3 \to \RRR^3$ that map any given $N$ points to any other given $N$ points. The \emph{distances} between the points carry the relevant information about, for example, what is written in a newspaper.
\end{itemize}

\subsubsection{Evolving Geometry of Configuration Space}

This suggests to introduce, in addition to photons and gravitons, a dynamical metric. In fact, item (i) asks for a different kind of metric than item (iii): not a metric $g_{\mu\nu}$ on space-time $\M$, but instead a metric $\g_{\sigma\tau}$ on configuration-space-time $\cst$. What is that, configuration-space-time? It is the set on which the wave function is defined. In non-relativistic quantum mechanics of $N$ particles, it is $\cst_N = \RRR^{3N+1} = (\text{space})^N \times (\text{time})$; in relativistic quantum mechanics, we may take it to be $\cst_N = \M^N$; if we use a time foliation $\foliation$ in the space-time manifold $\M$, we may set $\cst_N = \bigcup_{\Sigma \in \foliation} \Sigma^N$, which is a $(3N+1)$-dimensional submanifold of $\M^N$. For identical particles, divide out the action of permutations (after subtracting the coincidence configurations). For a variable number $N$ of particles, form the union $\cst = \bigcup_{N=0}^\infty \cst_N$. 

We can form a metric $\g_{\sigma\tau}$ on $\cst_N$ once we have a metric $g_{\mu\nu}$ on space-time $\M$ by combining $N$ copies of $g_{\mu\nu}$ when forming $\M^N$ (and then, if we use a time foliation, restricting the metric from $\M^N$ to the submanifold $\cst_N$). But I propose to do the opposite: Obtain $g_{\mu\nu}$ from $\g_{\sigma\tau}$ and treat $\g_{\sigma\tau}$ as an independent variable governed by a law of its own. The rule for obtaining $g$ from $\g$ is to insert the actual configuration,
\begin{equation}\label{gG1}
  g(x) = \g(Q \cup x)\,.
\end{equation}
I will make this precise in a moment. Before, I remark that this construction works only if we allow a variable number of particles, if ``particles are just points''  (see the end of Section~\ref{sec:POVM}), and if we use a time foliation. We thus set 
\begin{equation}\label{cstdef}
  \Gamma(\Sigma) = \{q \subset \Sigma: \# q < \infty\} \,, \quad
  \cst = \bigcup_{\Sigma \in \foliation} \Gamma(\Sigma) \,.
\end{equation}
The set $\Gamma(\Sigma)$ can be regarded as the configuration space at ``time'' $\Sigma$. Together, these sets form a foliation
\begin{equation}\label{Foliationdef}
  \Foliation = \{\Gamma(\Sigma): \Sigma \in \foliation\}
\end{equation}
of $\cst$, which I will also call the ``time foliation.'' 

To make precise equation \eqref{gG1}, we associate with every time leaf $\Sigma \in \foliation$ the actual configuration $Q_\Sigma \in \Gamma(\Sigma)$, and first define the Riemannian metric on $\Sigma$ as the metric that the next particle would see. That is, if $x \in \Sigma$ and $u,v \in T_x\Sigma$ (the tangent space at $x$) then
\begin{equation}
  g_{\mu\nu}(x)\, u^\mu v^\nu = \g_{\sigma\tau}(Q_\Sigma \cup x)\,
  \tilde{u}^\sigma \tilde{v}^\tau
\end{equation}
with $\tilde u, \tilde v \in T_{Q_\Sigma \cup x} \Gamma(\Sigma)$ the appropriate lifts of $u,v$. To define the timelike and mixed components of $g_{\mu\nu}$, we introduce the vector field $\n^\sigma$ on $\cst$ as the unit normal vector field of the foliation $\Foliation$; that is, $\n^\sigma$ is orthogonal, relative to $\g_{\sigma\tau}$, on $\Gamma(\Sigma)$. Since
\begin{equation}
  T_{Q_\Sigma \cup x} \cst \subseteq T_x \M \oplus \bigoplus_{y \in Q_\Sigma} T_y \M \,,
\end{equation}
we can consider the component of $\n^\sigma (Q_\Sigma \cup x)$ lying in $T_x \M$, multiply it by $\sqrt{\# Q_\Sigma +1}$, call the result $n^\mu(x)$, and define it to be the unit normal on $\Sigma$ relative to $g_{\mu\nu}$, i.e., $n^\mu n^\nu g_{\mu\nu} = 1$ and $u^\mu n^\nu g_{\mu\nu} = 0$ for all $u \in T_x \Sigma$. This completes the definition of $g_{\mu\nu}$. (The factor $\sqrt{\# Q_\Sigma +1}$ is supposed to compensate for the fact that a tangent vector in $\M^N$ obtained by combining $N$ unit tangent vectors in $\M$ has length $\sqrt{N}$.)\footnote{Moreover, a lift $\tilde{u}\in T_{q\cup x} \cst$ is now defined for \emph{all} tangent vectors $u\in T_x\M$, not only those tangent to the time leaf $\Sigma$, if we use $\n^\sigma$ and $n^\mu$: We start with writing $u^\mu = v^\mu + s\, n^\mu(x)$, with $v^\mu$ the projection of $u^\mu$ to $T_x \Sigma$ and $s= u^\mu\, n_\mu$, and set $\tilde{u}^\sigma = \tilde{v}^\sigma + s \, \n^\sigma(q \cup x)$. Then the mapping $u \mapsto \tilde{u}$ is an isometry onto its image.}

Thus, a metric on configuration-space-time, together with particle trajectories and a time foliation, defines a metric on space-time, schematically
\begin{equation}
  \g_{\sigma\tau} + Q + \foliation \:\:\longrightarrow\:\: g_{\mu\nu} \,.
\end{equation}
But where does the metric $\g$ on configuration-space-time $\cst$ come from? I propose that it be generated by an evolution law of its own. 

The most obvious possibility seems to be the higher-dimensional analog of the Einstein field equation, i.e., the Einstein field equation on $\cst$:
\begin{equation}\label{Einstein}
  {}^\sharp\! R_{\sigma\tau} - \frac{1}{2} {}^\sharp\! R\,  \g_{\sigma\tau}
  =  \kappa\, \T_{\sigma\tau} \,,
\end{equation} 
where ${}^\sharp\! R_{\sigma\tau}$ is the Ricci tensor of $\g$, ${}^\sharp\! R$ its scalar curvature, $\kappa$ the gravitation constant, and the reader should keep in mind that $\g$ is a Lorentzian metric in the sense that on the $N$-particle sector $\cst_N$ of $\cst$, which is a $(3N+1)$-dimensional manifold, $\g$ has signature $(+---\cdots)$, with one timelike direction and $3N$ spacelike ones. I will specify the right hand side of \eqref{Einstein} below. It is true in any dimension, not just 4, that \eqref{Einstein} is an evolution equation of second order and determines (up to diffeomorphisms) the geometry on all of $\cst_N$ if one specifies, on a (suitable) initial hypersurface, say $\Gamma(\Sigma)$, a Riemannian metric $\g|_{\Gamma(\Sigma)}$ and the extrinsic curvature.\footnote{This I learned from Gerhard Huisken.} 

In addition, I propose the following equation governing the relation between $\g$ and $\Foliation$:
\begin{equation}\label{lawsharpF}
  {}^\sharp \nabla_\sigma \n_\tau - {}^\sharp \nabla_\tau \n_\sigma =0\,,
\end{equation}
where ${}^\sharp \nabla$ is the covariant (Levi-Civit\`a) derivative defined by the metric $\g$. This equation is the cousin of \eqref{lawF}. But whereas we regarded \eqref{lawF} as an evolution law for $\foliation$ given the metric $g_{\mu\nu}$, \eqref{lawsharpF} is better regarded as part of the evolution law for $\g_{\sigma\tau}$. Indeed, for constructing successively the objects of the model, one might first choose the manifold $\M$ (without metric) and an arbitrary foliation $\foliation$, obtain from this the manifold $\cst$ and the foliation $\Foliation$ by \eqref{cstdef} and \eqref{Foliationdef}, then solve \eqref{Einstein} and \eqref{lawsharpF} together to obtain $\g_{\sigma\tau}$. Eq.~\eqref{lawF} then follows from the definition of $g_{\mu\nu}$.\footnote{Here is how. For infinitesimally close time leaves $\Sigma$ and $\Sigma'$, the distance (defined by $\g$) between the lifts $\Gamma(\Sigma), \Gamma(\Sigma')$ is, by \eqref{lawsharpF}, constant along $\Gamma(\Sigma)$. Put differently, there is an infinitesimal number $ds$ so that for every $q \in \Gamma(\Sigma)$, $q + \n(q) \, ds$ lies on $\Gamma(\Sigma')$. Therefore, for arbitrary but fixed $Q_\Sigma \in \Gamma(\Sigma)$ and all $x \in \Sigma$, by setting $q= Q_\Sigma \cup x$ we obtain that $x + n(x)\, (\#Q_\Sigma +1)^{-1/2}  \,ds$, being the $x$ component of $q+ \n(q) \, ds$, lies on $\Sigma'$. Thus, the distance (defined by $g$) between $\Sigma$ and $\Sigma'$ is constant along $\Sigma$, which is what \eqref{lawF} expresses.} (Then again, equations \eqref{Einstein} and \eqref{lawsharpF} may well put topological constraints on the possible choices for $\M$ and $\foliation$.) Since \eqref{lawsharpF} expresses that the timelike distance (defined by $\g$) between two nearby time leaves $\Gamma(\Sigma)$ and $\Gamma(\Sigma')$ is constant over every connected subset of $\Gamma(\Sigma)$, but since the manifold $\Gamma(\Sigma)$ is not connected, it is a natural idea to make the law a bit stronger than \eqref{lawsharpF} by postulating that it the distance between $\Gamma(\Sigma)$ and $\Gamma(\Sigma')$ globally constant, so that the constant is the same in every sector $\cst_N$. 

The source term $\T$ should consist of two contributions, $\T=\T_\text{particles} + \T_\text{e.m.}$ (with e.m.\ = electromagnetism), or more for further gauge fields. I could imagine that a reasonable particle term could be something like 
\begin{equation}\label{Tparticles}
  (\T_\text{particles})_{\sigma \tau} = \n_\sigma \, \n_\tau 
  \sum_{1\leq i<j\leq N} m_i m_j \delta(x_i-x_j)\,.
\end{equation}
This source term is concentrated on the coincidence configurations. Furthermore,
\begin{equation}\label{Fenergy}
  (\T_\text{e.m.})_{\sigma\tau} = -%\frac{1}{8\pi} 
  \F_{\sigma\rho} \, \F_{\tau}^{\:\:\rho} - *\F_{\sigma\rho} \; {*}\F_{\tau}^{\:\:\rho} \,,
\end{equation}
where $*$ denotes the Hodge operator and $\F_{\sigma\tau}$ is a 2-form on $\cst$, the curvature (or exterior derivative) of a $U(1)$ gauge connection (which I write as a 1-form $\A_\sigma$) on $\cst$. The expression \eqref{Fenergy} is literally the same formula as for the stress-energy tensor of the classical Maxwell field; but now the tensors live on $\cst$. Since the Dirac equation on $\cst$ requires a metric $\g_{\sigma \tau}$ on $\cst$ and a $U(1)$ gauge connection $\A_\sigma$ on $\cst$, it seems natural to treat both $\g_{\sigma\tau}$ and $\A_\sigma$ on an equal footing. We thus have three objects living on $\cst$:
\begin{equation}
  \g_{\sigma\tau} , \A_\sigma, \Psi\,.
\end{equation}
And $\A_\sigma$ should have its own evolution law, too. The simplest such law seems to be the higher-dimensional analog of the Maxwell field equations, i.e., the Maxwell equations on $\cst$:
\begin{equation}\label{Maxwell}
  d\F = 0, \quad {}^\sharp \nabla_{\sigma} \F^{\sigma\tau} = 4\pi \J^{\tau} \,,
\end{equation}
where $d$ means exterior derivative, and $\J$ is the source term. In classical electrodynamics, it would be the charge current density vector. In our case, I could imagine that a reasonable source term could be something like 
\begin{equation}\label{J}
  \J_\sigma = \n_\sigma \sum_{1\leq i<j\leq N} q_i q_j \delta(x_i-x_j)
\end{equation}
with $q_i$ the charge of particle $i$. Unlike in classical electrodynamics, this source term involves the \emph{product} of the charges.

We deal here with a kind of fields $\g,\A$ on configuration space. They are different from fields on space, such as classical fields, and different from quantum fields, which are operator-valued fields on space. Fields on configuration space are a bit like wave functions, as wave functions, too, are functions on configuration space. But they resemble more the potential $V$ of classical and quantum mechanics. A big difference is that potentials are usually regarded as fixed functions (think of the Coulomb potential on the configuration space of $N$ particles) and not as functions obtained by solving a differential equation. Thus, $\g$ and $\A$ are better thought of as \emph{evolving potentials} than as ``fields'', a word that would suggest something similar to either classical or quantum fields.

\subsubsection{Kiessling}

A pioneer of the idea of evolving potentials is Michael Kiessling. To my knowledge, he was the first to consider evolution equations for potentials on configuration space. In his two-part work \cite{Kie04a,Kie04b}, he attacks several problems at the same time: He deals with the ultraviolet divergence of classical and quantum electrodynamics (using the Born--Infeld equations instead of the Maxwell equations) and suggests steps towards extending Bohmian mechanics to QED, while introducing evolving potentials on configuration space and making the equations as relativistic as possible.\footnote{He also uses a time foliation (in his case given simply by some Lorentz frame) but tries to get on without it as long as possible.} I have borrowed the notation $\A$ from him, even though in his model, ``$\A$'' denotes something slightly different. His $\A$ corresponds, said somewhat simplified, to a potential on 3-space that depends on the electron configuration, in the sense that it is a function on $\RRR^{3} \times \RRR^{3N}$ in a setting with $N$ electrons. This is different from what I described above, where the $\A_\sigma$ field was a function on configuration space, corresponding to a function on $\RRR^{3N}$ in a setting with $N$ electrons. Kiessling also has such a function on configuration space $\RRR^{3N}$, which is needed as the gauge connection for defining the evolution of the wave function; he calls it $\tilde{A}$ and constructs it from his $\A$ function by inserting the actual position of an electron into the first slot. I have chosen here the somewhat simpler possibility of postulating directly the kind of field needed for the Dirac equation: a one-form $\A_\sigma$ on configuration space.

\subsubsection{On the Structure of the Model}

Note that the evolution of $\g$ and $\A$ does not depend on the actual particle configuration, and not on the wave function. Thus, the model I am presenting has a three-level hierarchical structure:
\begin{equation}
  \g,\A \:\:\longrightarrow \:\:
  \Psi\:\: \longrightarrow\:\:
  Q, g \,.
\end{equation}
The metric and gauge connection influence but are not influenced by the wave function, which influences but is not influenced by the particle trajectories. The metric $g$ is a function of $\g$ and $Q$. It has sometimes been objected to Bohmian mechanics that some principle of action and reaction be violated if the evolution of the wave function does not depend on the actual configuration. Here we encounter the same situation twice! I feel this makes the theory elegant and simple.

Electromagnetism has a \emph{dual structure} in this theory, consisting partly of the evolving potential $\A_\sigma$ on configuration space and partly of photons. This is surprising since classically, there is only one object, the vector potential $A_\mu$. Gravity has even a three-part structure in this theory: the metric $\g_{\sigma\tau}$ on configuration-space-time, the graviton particles, and the metric $g_{\mu\nu}$ on space-time, while classically, there is only $g_{\mu\nu}$.

As the output of the theory (its primitive ontology) I regard the triple
\begin{equation}\label{MgQ}
  (\M,g,Q)\,,
\end{equation}
where $Q = \bigcup_{\Sigma \in \foliation} Q_\Sigma$ is the set of all space-time points through which a particle passes. Thus, \eqref{MgQ} is a Lorentzian manifold with world lines on it. This is what has to be compared to the real world. All other variables, $\g_{\sigma\tau}, \A_\sigma, \foliation, \Psi$ had merely the role of generating this output.

Note that two triples related by a diffeomorphism are to be regarded as physically equivalent. Thus, strictly speaking, the output of the theory is a diffeomorphism class of triples \eqref{MgQ}. In the discussion so far, I have treated the manifold $\M$ as if it was given and fixed, but this should not be taken too seriously. I imagine that this attitude could be relaxed in favor of one regarding $\M$ itself as determined by the evolution laws, such as \eqref{Einstein} and \eqref{gG1}.

\subsubsection{No Multi-Time Evolution}

The introduction of evolving potentials on configuration space has consequences for the nature of the wave function. Recall that a multi-time wave function needs several Schr\"odinger equations, one for each time coordinate, and that these equations are consistent only if the condition \eqref{multitconsistency} is satisfied. In the presence of potentials on configuration space, the multi-time evolution of (the $N$-particle sector of) the wave function is consistent if and only if the potentials (i.e., the metric $\g_{\sigma \tau}$ and the gauge connection $\A_\sigma$) factorize, i.e., if they are of product form, $\g = g^{(1)} \otimes \cdots \otimes g^{(N)}$ and $\A=A^{(1)} \otimes \cdots \otimes A^{(N)}$. This is generically not the case, and thus the wave function is defined only on $\cst$, for configurations that are simultaneous relative to the time foliation. For example, for $N$ distinguishable particles, $\psi$ is defined on $\bigcup_{\Sigma \in \foliation} \Sigma^N \subset \M^N$ but not on all of $\M^N$. In other words, for 3-surfaces $\Sigma$ that are not time leaves there need not be an answer to the question, ``What is the quantum state on $\Sigma$?''

As a consequence, the time foliation becomes relevant at an early stage of the definition of the theory. It is not merely needed for defining the Bohmian trajectories, but already for defining the evolution of the wave function.

\subsubsection{Comparison With QED} 

Let us consider the case in which the metric $\g_{\sigma\tau}$ is flat (the appropriate product of Minkowski metrics) and the foliation $\foliation$ is flat, too, i.e., consists of parallel 3-planes corresponding to one fixed Lorentz frame. Then a possible solution for \eqref{Maxwell} and \eqref{J} on $\cst_N$ is
\begin{equation}
  \A_0 = \sum_{1\leq i<j\leq N} \frac{q_i q_j}{|x_i-x_j|} \,, \quad 
  \A_{\sigma} = 0 \text{ for } \sigma \neq 0\,,
\end{equation}
the Coulomb potential. The model then has become a Bohmian version of QED in the Coulomb gauge. Let me explain. 

When quantizing the Maxwell equation (see, e.g., \cite{CDG}), it is recommendable, because some of the Maxwell equations are constraints, to split the degrees of freedom of the classical Maxwell field into the dynamical ones (the transversal field in the Coulomb gauge) and the fixed ones (the longitudinal field, in effect the Coulomb potential, in the Coulomb gauge), and then quantize only the dynamical ones. As a result, the quantized field corresponds to photons, while the Coulomb potential remains as a contribution to the Hamiltonian. The model I outlined agrees with that, as it contains, in addition to the photons, the Coulomb potential in the form of $\A_\sigma$. 

The field operators $\hat{A}_\mu(x)$, for $x \in \M$, of QED then should arise according to
\begin{equation}
  \hat{A}_\mu = \text{multiplication by $\A$} + \text{photon creation} + 
  \text{photon annihilation.}
\end{equation}
More precisely, for $x \in \Sigma$ and $u \in T_x \M$, 
\begin{equation}
  u^\mu \, \hat{A}_{\mu}(x) \, \Psi(q) = \tilde{u}^\sigma \, \A_\sigma(q \cup x) \, \Psi(q) 
  + u^\mu (a^\dagger_{\mu} (x) \, \Psi)(q) + u^\mu (a_{\mu}(x) \, \Psi)(q)
\end{equation}
for all $q \in \Gamma(\Sigma)$, where $\tilde{u}$ is the lift of $u$ to $T_{q\cup x}\cst$ and $a^\dagger_{\mu}(x)$ and $a_{\mu}(x)$ are the photon creation and annihilation operators in position representation (at location $x \in \Sigma$). This equation reflects the dual structure of electromagnetism in this model, consisting of (i)~the evolving potential $\A$ and (ii)~the photons.

\subsubsection{Comparison With Quantum Gravity}

Can one define operators $\hat{g}_{\mu\nu}(x)$, for $x\in\M$, from the model that could be regarded as representing the field operators of the gravity field? I imagine that the definition could be, schematically,
\begin{equation}
  \hat{g}_{\mu\nu} = \text{multiplication by $\g$} + \text{graviton creation} + 
  \text{graviton annihilation.}
\end{equation}
That is, for $x \in \Sigma$, one could define an operator acting on $\Hilbert_\Sigma$ by
\begin{equation}\label{hatgdef}
  \hat{g}_{\mu\nu}(x) \, \Psi(q) = \g(q \cup x) \, \Psi(q) + (a^\dagger_{\mu\nu} (x) \, 
  \Psi)(q) + (a_{\mu\nu}(x) \, \Psi)(q)
\end{equation}
for all $q \in \Gamma(\Sigma)$, where $a^\dagger_{\mu\nu}(x)$ and $a_{\mu\nu}(x)$ are the graviton creation and annihilation operators in position representation (at location $x \in \Sigma$), and $\g(q \cup x)$ is to be understood in the same way as in the construction of $g_{\mu\nu}$ from $\g_{\sigma\tau}$. It would be interesting to find out whether the model fits together in this way with known approaches to quantum gravity.

Conversely, we may start from a given quantum gravity theory and ask what data we need to construct a primitive ontology similar to the one of the present model, i.e., particle world lines and a metric $g_{\mu\nu}$. Here is my guess: Suppose we are given a manifold $\M$, a foliation $\foliation$, a Hilbert space $\Hilbert$, a state vector $\Psi \in \Hilbert$ (which is fixed in the Heisenberg picture), operators $\hat{g}_{\mu\nu}(x)$ acting on $\Hilbert$ for every $x\in\M$, and a POVM $\hat P_\Sigma$ on $\Gamma(\Sigma)$ acting on $\Hilbert$ for every $\Sigma \in \foliation$. (The time evolution is encoded, according to the Heisenberg picture, in the family of position POVMs $\hat P_\Sigma$.) Then the methods of Bell-type QFT should provide random trajectories $Q = (Q_\Sigma)_{\Sigma \in \foliation}$, and we could define
\begin{equation}
  g_{\mu\nu}(x) = \frac{\sp{\Psi}{\hat{P}_\Sigma(Q_\Sigma) \, \hat{g}_{\mu\nu}(x) 
  \, \hat{P}_\Sigma(Q_\Sigma)|\Psi}}{\sp{\Psi}{\hat{P}_\Sigma(Q_\Sigma)|\Psi}}
\end{equation}
with $\Sigma$ the time leaf containing $x$. (This formula is a kind of inversion of \eqref{hatgdef}. The multiplication operator by $\g(q\cup x)$ is recovered from $\hat{g}_{\mu\nu}(x)$ by taking its diagonal part in the position representation defined by $\hat{P}_\Sigma$.)

\subsubsection{A Technical Note on Spin Spaces}

Spin spaces normally carry a mathematical structure related to the space-time metric. In the Dirac formalism, this structure is represented by the gamma matrices $\gamma(x) \in \CCC T_x\M^* \otimes D_x \otimes D_x^*$ (where $*$ denotes the dual space and $D_x$ the complex-4-dimensional Dirac spin space) and related to the space-time metric $g_{\mu\nu}$ by
\begin{equation}\label{gammag}
  \gamma_\mu \gamma_\nu + \gamma_\nu \gamma_\mu = 2g_{\mu\nu}\,.
\end{equation}
Equivalently, in the two-spinor formalism \cite{PR84}, this structure is represented by an anti-symmetric bilinear form $\varepsilon_{AB}(x)$ on complex-2-dimensional spin-space $S_x$ and an isomorphism $\delta: S_x \otimes \overline{S}_x \to \CCC T_x \M$ (with $\overline{S}_x$ the complex conjugate space of $S_x$), which are related to the metric by
\begin{equation}\label{epsilong}
  \varepsilon_{AB} \, \bar{\varepsilon}_{A'B'} = 
  \delta^\mu_{AA'} \, \delta^\nu_{BB'} \, g_{\mu\nu} \,.
\end{equation}
Since this structure, either $\gamma_\mu$ or $\varepsilon_{AB}$ and $\delta^\mu_{AA'}$, is needed for writing down the Dirac equation and the wave equations for photons and for gravitons, we need to define from $\g$ corresponding structures $\sgamma$, $\sepsilon$, and $\sdelta$ on $\cst$.

On the $N$-particle sector $\cst_N$, the analogs of $\gamma_\mu$ and $\varepsilon_{AB}$ at $x=(x_1,\ldots,x_N)\in\cst_N$ are $\sgamma_{\mu_1\ldots \mu_N}$ and $\sepsilon_{A_1\ldots A_N,B_1\ldots B_N}$, defined on the spin space $D_{x_1} \otimes \cdots \otimes D_{x_N}$ respectively $S_{x_1} \otimes \cdots \otimes S_{x_N}$. For example, in case of a given background 4-metric $g_{\mu\nu}$ (with accompanying $\gamma_\mu$ and $\varepsilon_{AB}$ on $\M$) one would set (abbreviating $\mu_1, \ldots, \mu_N$ as $\vec{\mu}$ etc.)
\begin{equation}
  \sgamma_{\vec{\mu}}(x) = 
  \gamma_{\mu_1}(x_1) \otimes \cdots \otimes \gamma_{\mu_N}(x_N)\,,
\end{equation}
\begin{equation}
  \sepsilon_{\vec{A}\vec{B}}(x)= 
  \varepsilon_{A_1B_1}(x_1) \cdots \varepsilon_{A_NB_N}(x_N)\,,
\end{equation}
\begin{equation}
  \sdelta^{\vec{\mu}}_{\vec{A}\vec{A}'}(x) =
  \delta^{\mu_1}_{A_1A_1'}(x_1) \cdots \delta^{\mu_N}_{A_NA_N'}(x_N)\,.
\end{equation}
The obvious analogs of \eqref{gammag} and \eqref{epsilong} are the following relations between $\sgamma$, $\sepsilon$, $\sdelta$ and an object $\tilde{g}_{\vec{\mu}\vec{\nu}}$:
\begin{equation}\label{sgammadef}
  S( \sgamma_{\vec{\mu}} \sgamma_{\vec{\nu}}) =
   \tilde{g}_{\vec{\mu}\vec{\nu}}\,,
\end{equation}
where $S$ means symmetrization (so that the expression becomes symmetric in each pair of indices $\mu_i,\nu_i$), respectively
\begin{equation}\label{sepsilondef}
 \sepsilon_{\vec{A}\vec{B}} {}^\sharp \bar\varepsilon_{\vec{A}'\vec{B}'} = 
 \sdelta^{\vec{\mu}}_{\vec{A}\vec{A'}}\, \sdelta^{\vec{\nu}}_{\vec{B}\vec{B}'}
 \, \tilde{g}_{\vec{\mu}\vec{\nu}} \,.
\end{equation}
The object $\tilde{g}$ is a metric on the product space $T_{x_1}\M \otimes \cdots \otimes T_{x_N}\M$, whereas $\g$ is a metric on (the subspace $T_x\cst$ of) the direct sum $T_{x_1}\M \oplus \cdots \oplus T_{x_N}\M$. To obtain an object like $\tilde{g}$ from $\g$, the simplest rule I can think of is to set, for $u_i,v_i \in T_{x_i}\M$ and $\tilde{u}_i, \tilde{v}_i$ their lifts in $T_x\cst$,
\begin{equation}
  \tilde{g}_{\vec{\mu}\vec{\nu}} (x) \prod_{i=1}^N u_i^{\mu_i} 
  \, v_i^{\nu_i} = \prod_{i=1}^N \g_{\sigma \tau}(x_1,\ldots,x_N) \, \tilde{u}_i^{\sigma}
  \, \tilde{v}_i^\tau \,.
\end{equation}
Then $\sgamma$, $\sepsilon$, and $\sdelta$ can be regarded as defined by \eqref{sgammadef} and \eqref{sepsilondef}.

\subsection{The Physical Hilbert Space}

In order to make the Hamiltonian bounded from below (to avoid the catastrophic behavior that two interacting particles become faster and faster while their energies approach $\infty$ and $-\infty$, respectively), one has to restrict the Hilbert space. Thus, there are two Hilbert spaces to be considered, the \emph{extended Hilbert space} $\Hilbert_\ext$ which contains also the negative-energy states and the \emph{physical Hilbert space} $\Hilbert_\phys$ which contains only the physical states, roughly those with purely positive energy contributions. For example, for the Dirac equation of one particle in Minkowski space-time, $\Hilbert_\ext = L^2(\RRR^3,\CCC^4)$, and $\Hilbert_\phys$ is usually defined as the positive spectral subspace of the free Dirac Hamiltonian. This leads to the question how to define $\Hilbert_\phys$ in our model. Already in the case of a Dirac particle in a curved (non-stationary) background space-time geometry, there is, to my knowledge, no canonical, natural way of selecting $\Hilbert_\phys$.\footnote{Specifically, the following difficulties arise: The spectral gap between $-mc^2$ and $mc^2$, present for the free Dirac operator in Minkowski space-time, may disappear in curved space-time. To split the spectrum of the Dirac operator at zero seems arbitrary and is not gauge invariant. The fact that the free Dirac operator is concentrated on the (future and past) mass shell in Fourier space can no longer be exploited because Fourier transformation is not defined in a generic curved space-time.} 

I see two possibilities. First, it might be possible to use the time foliation for selecting $\Hilbert_\phys$. Second, we might give up the attempt at \emph{defining} $\Hilbert_\phys$ and to obtain it instead by \emph{evolving} it as part of the state description. In that case, we would specify $\Hilbert_\phys(t=0)$ at time $0$ as part of the initial data, regard it as part of the variables of the theory, and obtain $\Hilbert_\phys(t)$ by some evolution law.\footnote{The simplest such law that I could think of is this: Begin with introducing a time coordinate $t:\cst \to \RRR$ whose level sets are the time leaves $\Gamma(\Sigma) \in \Foliation$. Use $\g_{\sigma\tau}$ to form the vector field ${}^\sharp \nabla^\rho t/({}^\sharp \nabla^\sigma t \, \g_{\sigma\tau} \, {}^\sharp \nabla^\tau t)$ on $\cst$. Use the flow defined by this vector field for identifying $\Gamma(\Sigma(t))$ with $\Gamma(\Sigma(0))$, and the connection $\A_\sigma$ for identifying their bundles of spin spaces. This yields a linear operator $I_t:\Hilbert_{\ext,\Sigma(0)} \to \Hilbert_{\ext,\Sigma(t)}$, not necessarily unitary, and we could take $\Hilbert_\phys(t) = I_t(\Hilbert_\phys(0))$.}

In our case, $\Hilbert_\ext$ is, for every $\Sigma \in \foliation$, the tensor product of several Hilbert spaces, one for electrons, one for positrons, one for left-handed photons, one for right handed photons, and so on. The one for electrons, for example, is the subspace of anti-symmetric functions in $L^2(\widehat{\Gamma(\Sigma)}, B)$, where $\widehat{\Gamma(\Sigma)}$ is the universal covering space of $\Gamma(\Sigma)$ and $B$ is the appropriate bundle of spin spaces. The Hamiltonian $H$ then needs to include suitable (creation and annihilation) terms that keep the wave function $\Psi$ from leaving $\Hilbert_\phys$. For these terms I have no conrete proposal. Once these terms are specified, the theory has a four-level hierarchy:
\begin{equation}
  \g,\A \:\:\longrightarrow\:\: 
  \Hilbert_\phys, H \:\:\longrightarrow\:\: 
  \Psi \:\:\longrightarrow\:\: 
  Q,g \,. 
\end{equation}

\subsection{The Time Foliation}

Can one observe the time foliation? That is, can one determine experimentally which 3-surfaces the time leaves are? If $\foliation = \foliation_\mathrm{BB}$ (the foliation defined in Section~\ref{sec:foliation} of the surfaces of constant distance from the Big Bang), then of course one can, by determining the age of the universe at every point. (Moreover, $\foliation_\mathrm{BB}$ may coincide with the rest frame of the cosmic microwave background radiation.) But this has nothing to do with quantum theory, and thus should not count as a serious observation of the time foliation. It seems that a serious observation of $\foliation$ should constitute a violation of relativistic covariance. This suggests that $\foliation$ be unobservable. But presumably the model I have presented entails that there are quantum experiments observing the time foliation, as I see no reason in the model why it should be unobservable. It would be interesting to think up an experiment for which the model predicts that its result reveals the time foliation.

\section*{Acknowledgments}

I thank Detlef D\"urr (M\"unchen), Shelly Goldstein (Rutgers), Gerhard Huisken (Potsdam and T\"ubingen), Michael Kiessling (Rutgers), Frank Loose (T\"ubingen), James Taylor (Baltimore), Stefan Teufel (T\"ubingen), and Nino Zangh\`\i\ (Genoa) for helpful discussions at various times on various topics related to this article.

\end{document}